\documentstyle[preprint,aps]{revtex}
\newcommand{\klugl}{\makebox{\raisebox{-3pt}
{$\,\stackrel{\scriptstyle{<}}{\scriptstyle{\sim}}\,$}}}
\newcommand{\grugl}{\makebox{\raisebox{-3pt}
{$\,\stackrel{\scriptstyle{>}}{\scriptstyle{\sim}}\,$}}}
\newcommand{\eqn}[1]{(\ref{#1})}
\newcommand{\Fig}[1]{Figure~\ref{#1}}
\newcommand{\fig}[1]{fig.~\ref{#1}}
\newcommand{\ek}[1]{\epsilon_{#1}}
\begin{document}
\draft
\title{Raman Scattering from Frustrated Quantum Spin Chains}
\author{Wolfram Brenig}
\address{Institut f\"ur Theoretische Physik, Universit\"at zu K\"oln
Z\"ulpicher Str. 77, 50937 K\"oln, Germany}
\date{\today}
\maketitle
\begin{abstract}
Magnetic Raman scattering from a frustrated spin--1/2 Heisenberg--chain
is considered with a focus on the uniform phase of the spin--Peierls
compound CuGeO$_3$. The Raman intensity is analyzed in terms of a
Loudon--Fleury scattering process using a spinless--fermion
mean--field theory developed for the frustrated spin--chain. A
comparison to experimental data is presented and the frustration and
temperature dependence is studied.  In good agreement with observed
spectra a broad inelastic four--spinon continuum is found at low
temperatures above the spin--Peierls transition.  At
high temperatures the intensity develops a quasi--elastic line
analogous to experiment.
\end{abstract}

\section{Introduction}
Elementary excitations of one--dimensional (1D) quantum spin--chains
exhibit a number of remarkable subtleties, such as the magnon
continuum of the isotropic 1D spin--1/2 Heisenberg model or the
excitation--gap for integer--spin systems as conjectured by
Haldane. Recent and extensive studies of 'spin--ladder' compounds as
well as of the germanate CuGeO$_3$ \cite{Hase93} have lead to renewed
interest in quantum spin--chains.  In this context, magnetic Raman
scattering, if symmetry--allowed, has developed into an important tool
to investigate the local spin dynamics
\cite{Kuroe94,Loosdrecht96,Lemmens96}. Here I will establish a simple
framework to interpret the magnetic Raman scattering from a frustrated
quantum spin--chain with a particular focus on the uniform phase of
CuGeO$_3$.

CuGeO$_3$ is a quasi--1D anorganic spin--Peierls compound with a
dimerization transition at a temperature $T_{SP}\simeq 14K$
\cite{Hase93,Nishi94,Pouget94}.  Its structure comprises of weakly
coupled CuO$_2$ chains, with copper in a spin--1/2 state
\cite{Voellenkle67}.  Since the nearest--neighbor ($n.n.$)
exchange--coupling between copper spins along the CuO$_2$ chains is
strongly reduced by almost orthogonal intermediate oxygen states
\cite{Braden96} the next--nearest--neighbor ($n.n.n.$) exchange is
relevant. Both, $n.n.$ and $n.n.n.$ exchange, are antiferromagnetic
\cite{Geertsma96} implying intra--chain frustration. A minimal model
of CuGeO$_3$ is the $J_1$--$J_2$--$\delta$ model
\begin{equation}\label{1}
H = J_1 \sum_{l} [(1+(-)^l\delta) {\bf S}_l\cdot{\bf S}_{l+1} + \alpha
{\bf S}_l\cdot{\bf S}_{l+2} ]
\;\;\;.\end{equation}
Here, ${\bf S}_l$ is a spin--1/2 operator, $J_1\approx 160K$ 
\cite{Castilla95,Riera95,Buechner96} is the
$n.n.$ exchange--coupling constant and $\delta$ resembles the lattice
dimerization which is finite for $T<T_{SP}$ only. $\alpha$ is the
intra--chain frustration--ratio $\alpha=J_2/J_1$ where $J_2$ is the
$n.n.n.$ exchange--coupling constant. A final consensus on the precise
magnitude of $\alpha$ is still lacking. Studies of the magnetic
susceptibility have resulted in $\alpha\approx 0.24$ \cite{Castilla95}
as well as in $\alpha\approx 0.35$ \cite{Riera95}. The latter is
consistent with a comparative investigation of magnetic susceptibility
{\em and} thermal expansion \cite{Buechner96}.  Therefore, very
likely, CuGeO$_3$ displays a frustration induced contribution to the
spin--gap at zero temperature irrespective of the actual lattice
dimerization \cite{Chitra95}. Depending on the magnitude of the
frustration the zero--temperature dimerization $\delta(T=0)$ has to be
on the order of $\delta(0) = 0.01 ... 0.03$ such as to enforce the
size of the spin--gap observed in inelastic neutron scattering (INS)
\cite{Nishi94}.

Magnetic excitations in CuGeO$_3$ are clearly distinct among the
uniform, i.e. $T>T_{SP}$, and the dimerized, i.e. $T<T_{SP}$,
phase. While the dynamic structure factor exhibits a gapless
two--spinon continuum similar to that of the 1D Heisenberg chain above
$T_{SP}$ \cite{Arai96}, well defined magnon--like excitations have
been observed below $T_{SP}$ \cite{Ain96}. These magnons reside within
the spin--gap and are split off from the two--spinon continuum.  They
have been interpreted as two--spinon bound states
\cite{Uhrig96a,Fledderjohann96}.

Magnetic Raman scattering from CuGeO$_3$ has been observed both, in
the dimerized as well as in the uniform phase
\cite{Kuroe94,Loosdrecht96,Lemmens96}.  In the low temperature uniform
phase, for $T_{SP}<T\ll J_1$, the Raman spectrum displays a broad
continuum centered at $\hbar\omega\approx 2.4J_1$. Early on this
continuum has been related to four--spinon excitations
\cite{Kuroe94,Loosdrecht96,Lemmens96,Muthukumar96}.  In the high
temperature uniform phase, and in addition to the four--spinon
continuum, the spectra show a pronounced quasi--elastic line. In the
dimerized phase the Raman intensity develops a gap at approximately
1.5--1.8$\Delta_{ST}$ \cite{Gros96a} where $\Delta_{ST}$ is the
singlet--triplet gap observed in INS \cite{Nishi94}.  Moreover, four
characteristic peaks appear in the spectrum the lowest one of which at
30cm$^{-1}$ 'coincides' with the Raman gap. At present the
interpretations of these peaks are controversial. Tentatively the
30cm$^{-1}$ line has been attributed to a {\em continuum} of
two--magnon bound states \cite{Loosdrecht97,Gros96a}. A possible dimensional
crossover effect has been invoked for the 225cm$^{-1}$ line
\cite{Loosdrecht96,Gros96a}. In this respect the influence of
inter--chain coupling in CuGeO$_3$ is an open problem \cite{Uhrig96b}.

Here I will focus on Raman scattering from the uniform phase of the
$J_1$--$J_2$--$\delta$ model, i.e. at $\delta=0$. First I will
describe a spinless--fermion mean--field theory to treat the
$J_1$--$J_2$ Hamiltonian. Next the Raman intensity is expressed in
terms of a four--fermion correlation function.  Finally results for
the Raman spectra are compared with experimental findings and are
contrasted against other theoretical approaches.

\section{Mean--field theory of the $J_1$--$J_2$ model}
\label{II}
The spinless--fermion mean--field (MF) theory for the $J_1$--$J_2$
model is based on the Jordan--Wigner (JW) representation
\cite{Jordan28} of the spin algebra,
i.e. $S^z_l=(c^{\dagger}_lc^{\phantom{\dagger}}_l-\frac{1}{2})$ and
$S^+_l=(-)^l\prod_{j<l}(1-c^{\dagger}_jc^{\phantom{\dagger}}_j)
c^{\dagger}_l$, where $c^{(\dagger)}_l$ are fermion
operators. Inserting this into \eqn{1} at $\delta=0$ one obtains
\begin{eqnarray}\label{5}
H=\sum_l[&&-\frac{1}{2}c^{\dagger}_lc^{\phantom{\dagger}}_{l+1}
-\frac{1}{2}c^{\dagger}_{l+1}c^{\phantom{\dagger}}_l
-c^{\dagger}_lc^{\phantom{\dagger}}_l+c^{\dagger}_l
c^{\phantom{\dagger}}_l
c^{\dagger}_{l+1}c^{\phantom{\dagger}}_{l+1}+\frac{1}{4}
+\frac{\alpha}{2}c^{\dagger}_lc^{\phantom{\dagger}}_{l+2}
+\frac{\alpha}{2}
c^{\dagger}_{l+2}c^{\phantom{\dagger}}_l
\nonumber \\
&&+\alpha(-c^{\dagger}_lc^{\phantom{\dagger}}_{l+2}c^{\dagger}_{l+1}
c^{\phantom{\dagger}}_{l+1}
-c^{\dagger}_{l+1}c^{\phantom{\dagger}}_{l+1}c^{\dagger}_{l+2}
c^{\phantom{\dagger}}_l
-c^{\dagger}_lc^{\phantom{\dagger}}_l+c^{\dagger}_l
c^{\phantom{\dagger}}_l
c^{\dagger}_{l+2}c^{\phantom{\dagger}}_{l+2}
+\frac{1}{4})]
\;\;\;,\end{eqnarray}
where $J_1$ is set to unity in section \ref{II} and \ref{III}.  $l$
runs over the lattice sites. The terms proportional to $\alpha$ are
absent in the JW representation of the isotropic $n.n.$ Heisenberg
model. In contrast to the $n.n.$ model, both, the transverse as well
as the longitudinal $n.n.n.$ exchange--interaction lead to
four--fermion vertices.  Longer range spin--exchange leads to even
higher order couplings, as is obvious from the JW representation which
implies $2l$--fermion vertices for $l$--th.--nearest--neighbor
spin--exchange. To proceed I treat Hamiltonian \eqn{5} in
MF--approximation. Allowing for all contractions of type $\langle
c^{\dagger}_lc^{\phantom{\dagger}}_m\rangle$ one gets
\begin{eqnarray}\label{7}
H_{MF}=\sum_k \{ &&
-[1+(A+B)(1-2\alpha)]\cos (k)\,c^{\dagger}_k c^{\phantom{\dagger}}_k
\nonumber \\
&& +\frac{i}{2}(A-B)(1+2\alpha)\sin (k)
(c^{\dagger}_k c^{\phantom{\dagger}}_{k+\pi} - c^{\dagger}_{k+\pi} 
c^{\phantom{\dagger}}_k) \} + const.
\;\;\;,\end{eqnarray}
where $k$ is the momentum, $A=A^{\star}=\langle c^{\dagger}_{2l}
c^{\phantom{\dagger}}_{2l+1}\rangle$, and $B=B^{\star}=\langle
c^{\dagger}_{2l+1} c^{\phantom{\dagger}}_{2l}\rangle$. In principle
contractions of type $D=D^{\star}=\langle
c^{\dagger}_{2l+2}c^{\phantom{\dagger}}_{2l}\rangle$ do occur,
however, their selfconsistent value can be shown to vanish
identically. Note that \eqn{7} allows for both, a uniform, if $A-B=0$,
and a gaped phase, if $A-B\neq 0$.  Here no attempt will be made to
describe the spin--dimer state using $A\neq B$ and the relevance of
the gaped solution of the MF--theory will be discussed elsewhere. In
the uniform case \eqn{7} resembles a single--band spinon gas with a
hopping amplitude $t(T,\alpha)$ to be determined selfconsistently
\begin{eqnarray}\label{8}
&&H_{MF}=\sum_k\ek{k} c^{\dagger}_k c^{\phantom{\dagger}}_k
\nonumber \\
&&\ek{k}=-t(T,\alpha)\cos (k)=-[1+2A(1-2\alpha)]\cos (k)
\\
&&A=\frac{2}{N}\sum_{0\leq k\leq\pi}\cos(k) f(\ek{k})
\nonumber
\;\;\;,\end{eqnarray}
where $f(\epsilon)=[\exp (\epsilon/T)+1]^{-1}$ is the Fermi
function. At zero temperature the spinon dispersion simplifies to
\begin{equation}\label{8a}
\ek{k}(T=0)=-[1+2(1-2\alpha)/\pi]\cos (k)
\;\;\;.\end{equation}
For vanishing frustration this is identical to Bulaevskii's result
$-(1+2/\pi)\cos (k)$ \cite{Bulaevskii62}. The latter is known to
compare reasonably well with the exact spinon dispersion
$\epsilon_k=-\pi/2\cos(k)$ by des Cloizeaux and Pearson
\cite{Cloizeaux62}.  For finite $\alpha$ the MF--theory results in a
frustration induced {\em softening} of the spinon stiffness which, for
$T=0$, can be expressed as
\begin{equation}\label{8c}
\frac{v_s(\alpha)}{v_s(0)}=1- \frac{4}{2+\pi} \alpha
\approx 1 - 0.778 \alpha
\;\;\;,\end{equation}
where $v_s(\alpha)=\partial\ek{k}/\partial k|_{k=\pi/2}$ is the spinon
velocity. Eqn. \eqn{8c} is qualitatively consistent with a numerical study
where $v_s(\alpha)/ v_s(0)\approx 1-1.12\alpha$ has been found
\cite{Fledderjohann96}.  In \fig{1} $t(T, \alpha)$ is depicted as a
function of temperature for various values of frustration. As is
obvious the MF spinon--stiffness is decreased, both, as a function of
increasing frustration and temperature. At zero temperature the MF
ground state energy is given by
$E_0(\alpha)=N(-1/\pi-1/\pi^2+2\alpha/\pi^2)$ which, for $\alpha=0$,
leads to $E_0(0)\approx -0.420$ \cite{Bulaevskii62}. This agrees
reasonably well with the Bethe--Ansatz result $E_0=1/4-\ln(2)\approx
-0.443$.  Moreover, the linear frustration dependence of the MF ground
state energy, $(E_0(\alpha)-E_0(0))/N\approx 0.203\alpha$ is close to
that found in finite--chain diagonalization \cite{Tonegawa87}, where
$(E_0(\alpha)- E_0(0))/N\approx 0.177\alpha$ for $\alpha\klugl 0.3$.

\section{Raman scattering}
\label{III}
The MF--theory is a convenient tool to study Raman scattering from
the $J_1$--$J_2$ model. The Raman vertex is given by Loudon--Fleury's
photon--induced super--exchange operator \cite{Fleury68}
\begin{equation}\label{2}
R=\sum_{lm} T_{lm}
({\bf E}_{in}\cdot {\bf n}_{lm})
({\bf E}_{out}\cdot {\bf n}_{lm})
{\bf S}_l\cdot {\bf S}_{m}
\;\;\;.\end{equation}
Here $T_{lm}$ sets the coupling strength, ${\bf E}_{in}$ (${\bf
E}_{out}$) refers to the field of the in(out)going light, and ${\bf
n}_{lm}$ labels a unit vector connecting the sites $l$ and $m$. By
symmetry, in a strictly 1D situation, \eqn{2} leads to scattering 
only for parallel polarization
along the c--directed chains. Since the real--space decay of $T_{lm}$
is comparable to that of the exchange integrals in \eqn{1} it is
sufficient to consider a Raman operator with at most $n.n.n.$
spin--exchange
\begin{equation}\label{3}
R=C \sum_{l}
({\bf S}_l\cdot{\bf S}_{l+1} + \beta {\bf S}_l\cdot{\bf S}_{l+2})
\;\;\;.\end{equation}
The scattering intensity $I(\omega)$ is obtained via the fluctuation
dissipation theorem $I(\omega)=\chi''(\omega)/(1-e^{-\omega/T})$ from
the dynamical susceptibility of the Raman operator
\begin{equation}\label{3a}
\chi''(\omega)=Im[\chi(\omega+i\eta)]=
Re\int^{+\infty}_0dt\; e^{i(\omega+i\eta)t}
\langle [ R(t), R ] \rangle
\;\;\;,\end{equation}
where $\langle ...\rangle$ denotes the thermal average.

Frustration of {\em unequal} magnitude regarding the $J_1$--$J_2$
model and the Raman operator is {\em mandatory} for non--vanishing
inelastic scattering.  If $\alpha = \beta$ the Hamiltonian for
$\delta=0$ and the Raman operator commute which leads to elastic
scattering only. The inelastic intensity which results from \eqn{3} is
identical to that of a Raman operator $R' = R -\gamma H$. Setting
$\gamma=C\beta/\alpha$ or $\gamma=C$ leads to scattering by a
renormalized $n.n.$ or $n.n.n.$ Raman operator only
\begin{equation}\label{4}
R_1 = C (1 - \frac{\beta}{\alpha}) \sum_{l} {\bf S}_l\cdot{\bf S}_{l+1}
\makebox[2cm][c]{or}
R_2 = C (\beta-\alpha) \sum_{l} {\bf S}_l\cdot{\bf S}_{l+2}
\;\;\;.\end{equation}
At present, an {\em exact} treatment of \eqn{3a} is not feasible and
the Raman intensity of any {\em approximate} evaluation will depend on
$\gamma$. In this respect $R_2$ is the proper choice for $\alpha,
\beta \ll 1$ since it guarantees that $I(\omega )\propto
(\beta-\alpha)^2$. This is not obvious for $R_1$ or other values of
$\gamma$. In those cases, even for $\alpha, \beta \rightarrow 0$, $R'$
may contain a component of order unity, proportional to the
Hamiltonian, which has to be projected out. To avoid this complication
$R_2$ will be considered hereafter.

Using the JW--fermions the Raman operator $R_2$ can be expressed in
terms of a four--fermion operator
\begin{eqnarray}\label{9}
&&R_2 = C (\beta-\alpha) \sum_{k,k',q} h(k,k',q)
c^{\dagger}_k c^{\phantom{\dagger}}_{k+q}c^{\dagger}_{k'} 
c^{\phantom{\dagger}}_{k'-q} + \Lambda_{1ph}
\\[.3cm] \label{9a}
&&h(k,k',q) = \cos (2q)-\cos (2k+q)-\cos (2k'-q)
\nonumber
\;\;\;,\end{eqnarray}
where $\Lambda_{1ph}$ labels all one--particle--hole excitations of the
JW--transform of $R_2$. Since these occur at zero total momentum they
do not contribute to inelastic Raman scattering within MF--theory
\cite{QPlifetime}. Using \eqn{9}, the Raman intensity \eqn{3a} is 
written in terms of the four--fermion propagator
\begin{eqnarray}\label{10}
\chi(\tau)=&& C^2 (\alpha-\beta)^2
\sum_{k,k',q,p,p',r} [ h(k,k',q) h(p,p',r)
\nonumber \\
&&\hskip3cm
\langle T_{\tau}(
c^{\dagger}_k(\tau) c^{\phantom{\dagger}}_{k+q}(\tau) 
c^{\dagger}_{k'}(\tau) c^{\phantom{\dagger}}_{k'-q}(\tau)
c^{\dagger}_{p'-r} c^{\phantom{\dagger}}_{p'} c^{\dagger}_{p+r} 
c^{\phantom{\dagger}}_p )\rangle ]
\;\;\;.\end{eqnarray}
Here $\tau$ is the imaginary time and $\chi''(\omega)$ results from
the usual analytic continuation
$\chi''(\omega)=-Im[\chi(i\omega_n\rightarrow \omega+i0^+)]$ where
$\omega_n=2n\pi T$ is a Bose Matsubara-frequency. On the level of
MF--theory the four--fermion propagator \eqn{10} is evaluated
neglecting all vertex corrections and using the MF one--particle
Green's functions corresponding to \eqn{8}. After a number of standard
manipulations I obtain
\begin{eqnarray}\label{11}
\chi''_{MF}(\omega)=&&-\frac{1}{2\pi^2} \int^{\pi}_{-\pi}dq \,
\int^{\pi}_{-\pi}dk \, \sum_{k'} \left\{
\frac{g^2(k,k'+q,q)}{\sqrt{[2t\sin (q/2)]^2-(\ek{k+q}-\ek{k}-\omega)^2}}
\right.
\nonumber \\  \nonumber \\
&& \left. [f(\ek{k})-f(\ek{k+q})] \,
[f(\ek{k'})-f(\ek{k'+q})]
[n(\ek{k'+q}-\ek{k'}+\omega)-n(\ek{k'+q}-\ek{k'})]
\phantom{\hskip-.5cm\frac{1^2}{1^2}}
\right\}
\;\;\;,\end{eqnarray}
where $g(k,k',q)=\frac{1}{2}[h(k,k',q)+h(k,k',k'-k-q)]$ and the discrete
sum on $k'$ runs over the set of solutions of
\begin{equation}\label{11a}
\sin(k'+q/2)=\frac{\ek{k+q}-\ek{k}-\omega}{2t\sin (q/2)}
\;\;\;; \makebox[.5cm]{} k'\in\;\; ]-\pi,\pi]
\;\;\;.\end{equation}
This concludes the MF--theory of Raman scattering.

\section{Discussion}
In \fig{fig2} I compare the MF--intensity obtained from a numerical
integration of \eqn{11} (solid and dashed lines) at $T=20K$ and
$12.8K$, for $J_1=160K$ and two values of frustration $\alpha=0.24$
and $\alpha=0.35$ with an experimental Raman spectrum observed at
$T=20K$ in the uniform phase of CuGeO$_3$\cite{Lemmens96,Muthukumar96}
(dashed--dotted line with markers). Phonon lines
at 184cm$^{-1}$ and 330cm$^{-1}$ have been removed. This Raman data is
consistent with other published results
\cite{Kuroe94,Loosdrecht96}. The absolute magnitude of the observed
Raman intensity as well as the coupling strength $C$ of the scattering
operator \eqn{3} are unknown quantities. Therefore the two sets of
MF--intensities for distinct $\alpha$ and the experimental spectrum
have been normalized on a scale of arbitrary units. The theoretical
parameters in \fig{fig2} represent no attempt at a 'best fit' to the
data, rather they have been chosen among those obtained independently
from studies of the magnetic susceptibility of CuGeO$_3$
\cite{Castilla95,Riera95,Buechner96}.  \Fig{fig2} displays reasonable
agreement between theory and experiment, although details of the
experimental spectrum, i.e. the structure at $375$ wave numbers and
the initial curvature are not reproduced by the MF--theory.  Evidently
a value of $\alpha=0.24$ \cite{Castilla95} is slightly more favored by
this comparison at $J_1=160K$ than $\alpha=0.35$
\cite{Riera95,Buechner96}.  Similar findings have been made in
numerical studies \cite{Gros96a}. However, since the maximum of the
MF--intensity roughly scales with the maximum of the
spinon--dispersion $\propto J_1 [1+2(1-2\alpha)/\pi]$, a better
agreement with $\alpha=0.35$ instead of $\alpha=0.24$ can also be
reached simply by increasing $J_1$ by $\sim$10\%.

The comparison in \fig{fig2} strongly corroborates a study
\cite{Muthukumar96} of Raman scattering from the uniform phase of
CuGeO$_3$ which is based the 'solitonic' mean--field description of
the XXZ Heisenberg chain by G\'omez--Santos \cite{Gomez90}. This
approach employs a domain--wall representation of the spin chain which
leads to formal developments very different from the straightforward
application of the Jordan--Wigner type of MF--theory presented here.

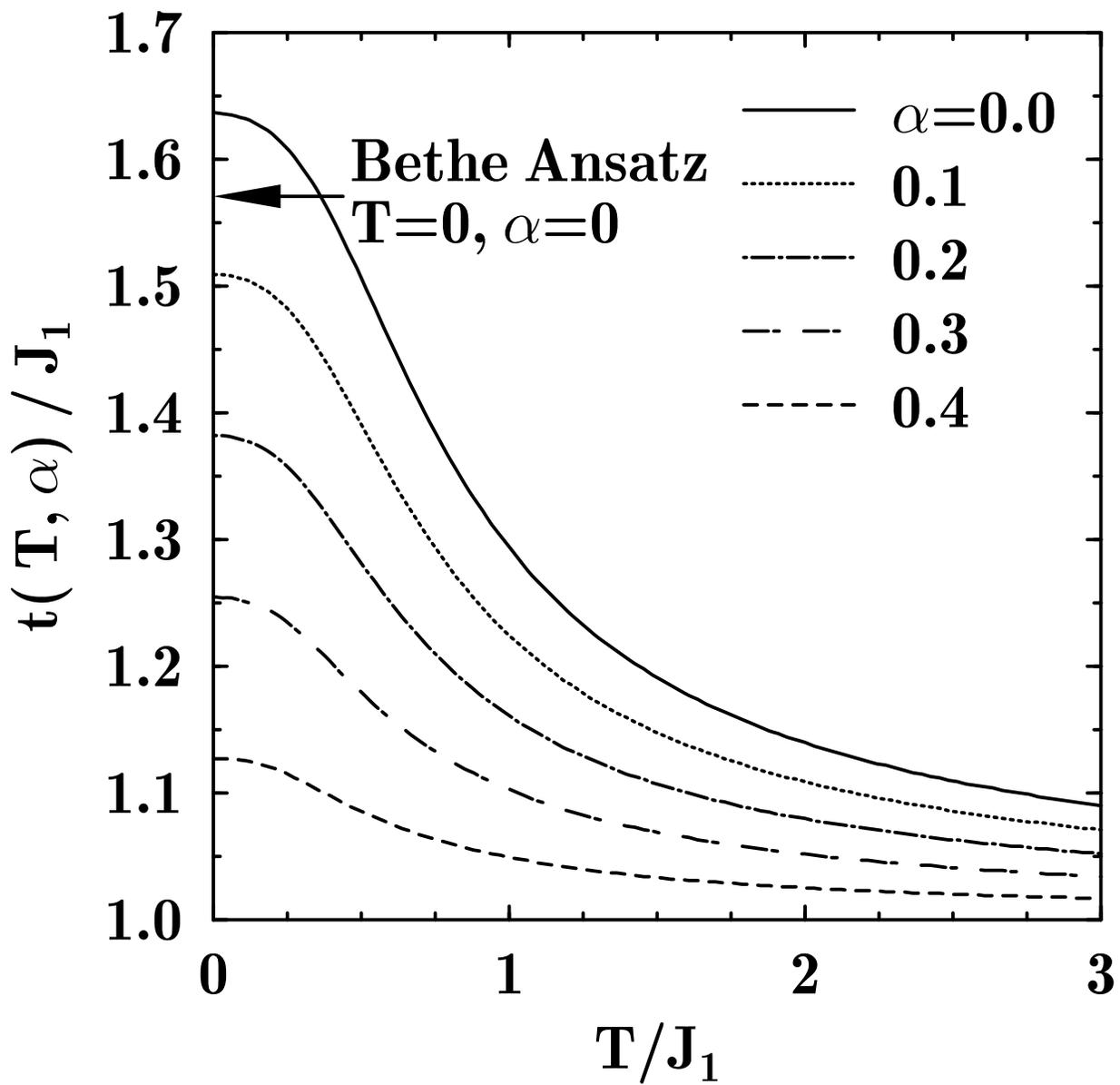
\begin{figure}
\caption[l]{
Temperature dependence of the mean--field hopping amplitude
for various values of frustration.
}
\label{fig1}
\end{figure}

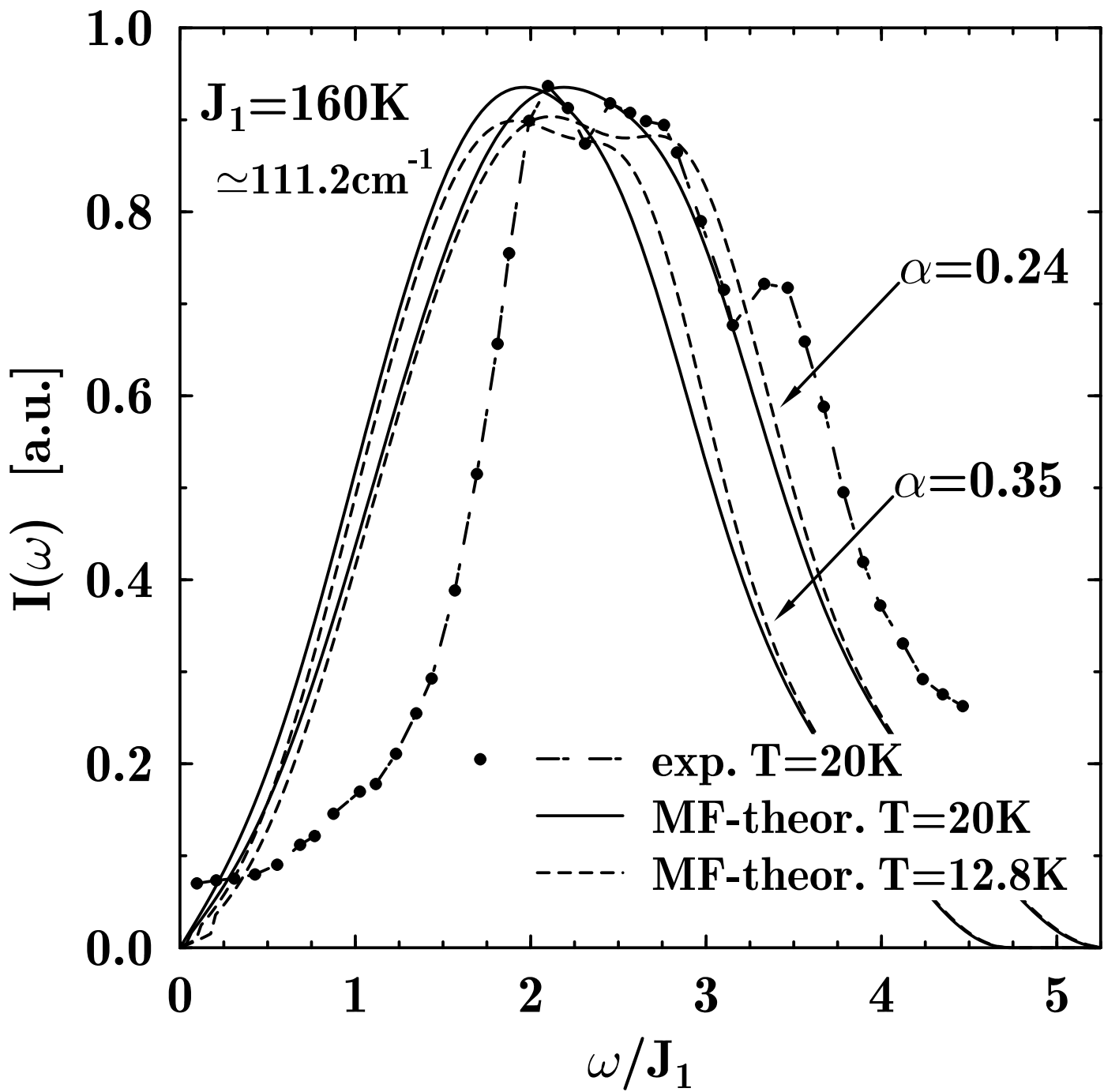
\begin{figure}
\caption[l]{
Mean--field Raman intensity (solid and dashed lines) at two
temperatures and for two values of frustration as compared to the
experimental Raman spectrum of CuGeO$_3$ \cite{Lemmens96,Muthukumar96}
(dashed--dotted line \& markers).}
\label{fig2}
\end{figure}

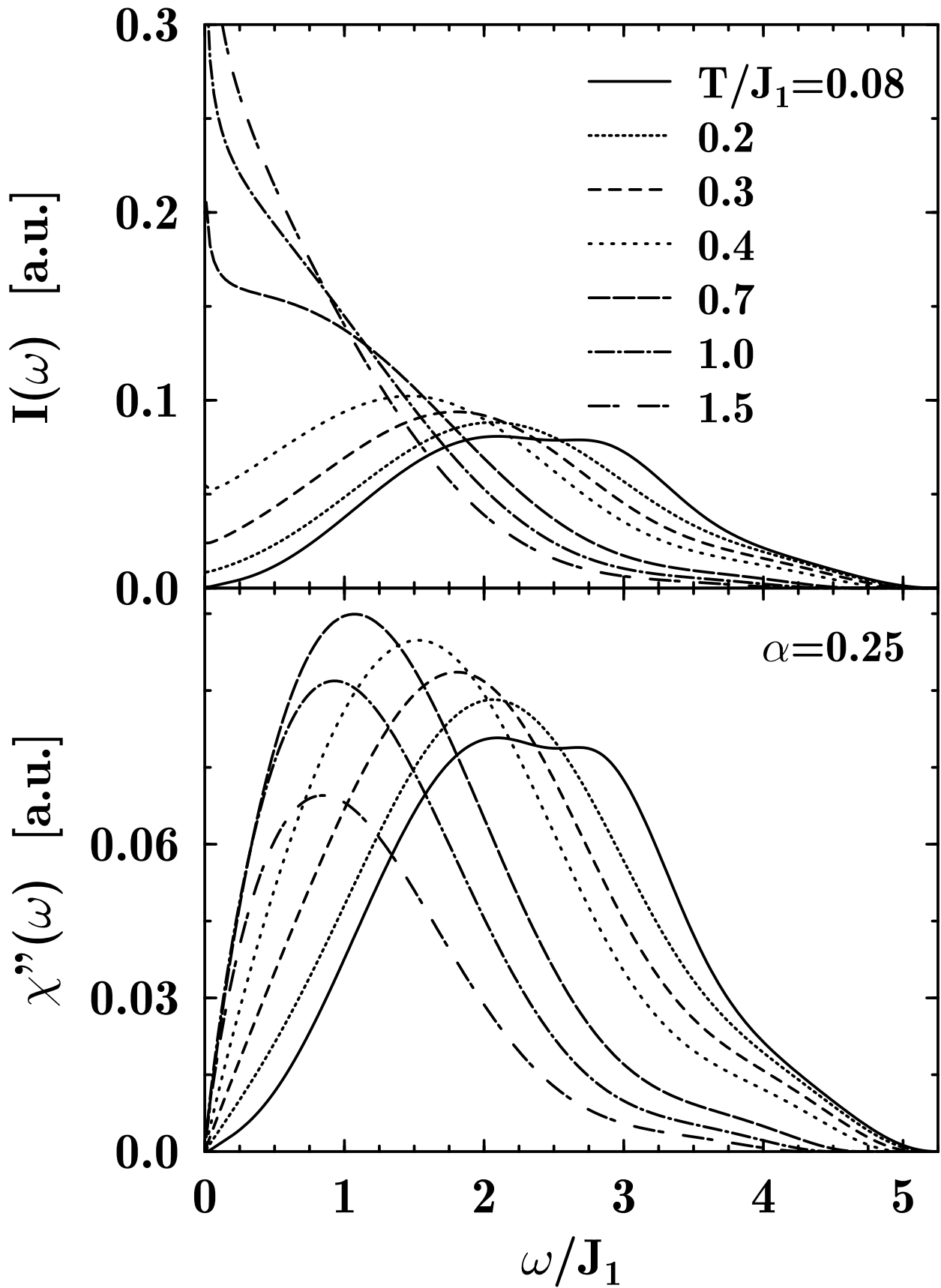
\begin{figure}
\caption[l]{
Mean--field Raman intensity and susceptibility for various temperatures.
}
\label{fig3}
\end{figure}

In \fig{fig3} the temperature dependence of the MF--spectrum is shown,
both, in terms of the intensity $I(\omega)$ -- which is observed in
experiment -- and the Raman--operator susceptibility $\chi''(\omega)$.
The latter exhibits a left--shift of its maximum upon increase of the
temperature. This leads to enhanced low--frequency spectral weight
which, by virtue of the Bose--prefactor, turns into a quasi--elastic
line in $I(\omega)$ for $T\grugl J_1$. A corresponding tendency has
been detected in finite--temperature Lanczos studies
\cite{Singh96}. In MF--theory the high--temperature enhancement of the
low--frequency spectral weight in $\chi''(\omega)$ is due to the
reduction of the spinon--stiffness as a function of increasing
temperature, see \fig{fig1}, and due to the 'smearing' induced by the
Fermi-- and Bose--functions contained in \eqn{11}. It is tempting to
relate the behavior depicted in \fig{fig3} to the high--temperature
quasi--elastic line observed in the uniform phase of CuGeO$_3$
\cite{Kuroe94,Loosdrecht96,Lemmens96}. However, in CuGeO$_3$ the
four--spinon continuum is only weakly shifted by temperature and
remains more clearly separated from the quasi--elastic line for
$T_{SP}\klugl T\klugl J_1$. This discrepancy may be due to an
overestimation of the temperature dependence of the spinon--stiffness
in MF--theory \cite{Fabricius96,Starykh97}.

Finally, I emphasize that the MF--theory gives only a limited
description of spinon interaction effects. In the case of the {\em
two}--spinon propagator this is known to result in an incorrect
description of the spectral weight distribution \cite{Mueller79}. This
caveat of MF--theory has stimulated a study of approximate vertex
corrections to the {\em four}--spinon--propagator \cite{Singh96}.
However, the resulting Raman spectra show no agreement with
experiment. At, present the impact of spinon interaction effects
beyond MF--theory on the Raman spectra remain unclear.

In conclusion I have described a finite temperature MF--theory for
frustrated spin--chains and consequently detailed its application to
magnetic Raman scattering.  At low--temperatures, in the gapless
phase, I find a scattering continuum which is due to frustration
induced four--spinon excitations and is compatible with observed Raman
spectra of CuGeO$_3$.  In the high--temperature regime the MF
Raman--intensity is dominated by a quasi--elastic line which results
from the temperature dependence of the spinon spectrum.

\section{Acknowledgments}
I am indebted to B. B\"uchner, P. Knoll, and E. M\"uller--Hartmann
for valuable discussions. Part of this work has been supported by
the Deutsche Forschungsgemeinschaft through the SFB 341.

\begin{thebibliography}{99}
\bibitem{QPlifetime}
In Raman scattering photon--momentum transfer to the electronic system
can be neglected which is implicit in the definition of $R$. Therefore
spinon one--particle--hole excitations will contribute to the
inelastic intensity only in case of a finite quasi--particle
relaxation rate, which is absent in MF--theory.
\end{thebibliography}
\end{document}